\documentclass [amssymb, twocolumn, amsmath, showpacs]{revtex4-1}
\usepackage{graphicx,epsfig,amsfonts,amssymb}
\usepackage{bm}
\usepackage{times}
\begin{document}

\title{Fluctuation Relations for  Current Components in Mesoscopic Electric Circuits}

\author {Sriram Ganeshan$^{1,3}$ and N.~A. {Sinitsyn}$^{2,3}$}
 \affiliation{$^1$Department of Physics and Astronomy, Stony Brook University, Stony Brook, NY 11794-3800, USA}
 \affiliation{$^2$Theoretical Division, Los Alamos National Laboratory, B213, Los Alamos, NM
87545}
 \affiliation{$^3$ New Mexico Consortium, Los Alamos, NM
87544, USA}

\date{\today}

\begin{abstract}
We present a new class of fluctuation relations, to which we will refer as {\it Fluctuation Relations for Current Components} (FRCCs). FRCCs can be used to estimate system parameters when  complete information about nonequilibrium many-body electron interactions is unavailable.  We show that FRCCs are often robust in the sense that they do not depend on some basic types of electron interactions and some quantum coherence effects.
 \end{abstract}


\date{\today}
 
\maketitle

\section{ Introduction}
 Fluctuation theorems \cite{FT-retro,FT-rev} are fundamental results in the nonequilibrium statistical mechanics. Their discovery led to optimism that they might serve as universal laws that had long been missing from the study of nonequilibrium systems. 

Fluctuation relations for currents \cite{FT-curr} play a special role in the physics of non-equilibrium transport. Measurements of statistics of heat production generally involve tracing complete system's stochastic trajectories. This is often an experimentally challenging problem. Measurements of currents are relatively simple. Formally, a current can be found just from the knowledge about an initial and a final state of a system, and no separation of the measured system from the heat bath is required. Moreover, experimental studies of fluctuation theorems typically need large statistics of events because any Gaussian distribution of any variable trivially satisfies a standard fluctuation relation. Hence, nontrivial fluctuation relations can be identified only if statistics of events are obtained beyond the domain of Gaussian fluctuations. Non-Gaussian statistics in nanoscale conductors is due to the shot noise of electrons at non-equilibrium conditions \cite{nazarov-book}. Recently, it became possible to experimentally study such non-Gaussian fluctuations in mesoscopic electric circuits \cite{noise-exp}. 

In application to an electric circuit with two lead contacts, the Fluctuation Theorem predicts that the probability distribution $P[q]$ of observing a charge $q$ passed between two lead contacts with a voltage difference $V$ satisfies the law \cite{FT-curr}:
\begin{equation}
P[q]/P[-q] = e^{  Fq },
\label{ft-1}
\end{equation}
where $F=V/k_{\rm B}T$. 
Here $T$ is the temperature,  and $k_{\rm B}$ is the Boltzmann's constant. 
Eq. (\ref{ft-1}) is expected to be universal, i.e. it should be valid independently of the type of electron interactions in a conductor. 
Surprisingly, recent experimental work \cite{FTIC-exp} has shown that the law (\ref{ft-1}) can fail in an electric circuit,  but could be salvaged under the experimental conditions of \cite{FTIC-exp}  if the parameter $F$ is suitably renormalized by a factor $\sim 10^{-1}$. 

The need to modify (\ref{ft-1}) was qualitatively explained in  \cite{FTIC-exp} by presence of a feedback between measured and measuring circuits. For example,  in experiment {\cite{FTIC-exp}, the nanoscale circuit  
was connected to a read-out circuit made of an additional tunnel junction, which was coupled to its own leads. When interactions between measured and measuring currents are involved, 
the  fluctuation theorem is applicable only to the {\it total system} that includes both the studied circuit and the measuring one but this does not imply (\ref{ft-1}) for a single current component any more. 

This argument only partly explains the experimental result in \cite{FTIC-exp} because it makes unclear why, after considerable renormalization of parameters, an individual current through a specific lead contact again satisfies (\ref{ft-1}).
For example, according to Crook's equality, the heat $W$, dissipated by a complete system, satisfies the relation 
\begin{equation}
P[W]/P[-W] = {\rm exp} \left( W/k_{\rm B} T \right).
\label{crooks}
\end{equation}
 We can introduce the vectors, ${\bm q}$ and ${\bm V}$, whose components are, respectively, the numbers of electrons passed through individual lead contacts, and individual voltages at corresponding leads. Numbers of vector components is equal to the number of independent lead contacts in the circuit.  The dissipated heat can then be expressed as 
$W={\bm V} {\bm \cdot} {\bm q}$, which corresponds to the standard fluctuation relation for multicomponent current \cite{FT-curr}:
\begin{equation}
P[{\bm q}]/P[-{\bm q}] = {\rm exp} \left( {\bm V }\cdot {\bm q} /k_{\rm B} T \right).
\label{ft-2}
\end{equation}
Apparently, a specific current component $q_k$ is not proportional to the dissipation function $W$ because the latter depends on all current components. 
Consequently, although the full vector of currents satisfies Eq.(\ref{ft-2}) there seems to be no reason why an individual current through a specific lead contact should satisfy a fluctuation relation (\ref{ft-1}) in a multicomponent system. 

 The appearance of fluctuation relations (\ref{ft-1}) for specific current components
 at renormalized values of parameters remains poorly understood. Recent studies \cite{gaspard-11} showed that {\it Fluctuation Relations for Current Components} (FRCC)s, i.e.  for currents through specific links  of a circuit, can appear in some limits of  a model that corresponds to a 4-state Markov chain kinetics. However, generalizations of this result have been unknown. Another possibility to explain experimentally observed FRCC was based on separation of time scales \cite{FTIC-exp}, i.e. if one current component is macroscopic in comparison to another one, the former can be considered as part of environment so that one can introduce effective dissipation function for the second, microscopic, current component.  Such an explanation, however, can be justified only in very specific limits and relatively simple circuit geometries.

In this article, we prove that there are, in fact, general conditions under which FRCCs should hold for currents through certain links in complex mesoscopic electric circuits coupled to multiple lead contacts.
Unlike standard fluctuation theorems, there is no direct relation between FRCCs and the  dissipated heat in a system. Instead, we will show that FRCCs follow from the observation that statistics of particle currents  depends only on probabilities of single particle {\it geometric trajectories} while the information about  time moments, at which particles make transitions along such trajectories, is irrelevant. 
Then, there can be purely topological constraints on contributions of  geometric trajectories to currents through some links of a circuit. For example,
if a link does not belong to any loop of a graph, and  if transitions through this link in opposite directions are counted with opposite signs, then any geometric trajectory can make only $\pm 1$ or $0$ valued currents through such a link. We will show  that this restricts statistics of currents through this link to satisfy Eq. (\ref{ft-1}) even when there are no detailed balance constraints on kinetic rates. 
Another interesting topological constraint appears when there is only one reservoir that supplies/absorbs particles to/from the system. Then all single particle geometric trajectories have to be cyclic, and one can make a correspondence between currents through links and independent cycles that a trajectory makes. Probabilities of independent cycles are known to satisfy relations of type (\ref{ft-1}) \cite{FT-curr}, which eventually results in FRCCs in such a circuit. 
We will also show that  FRCCs are robust against adding important many-body interactions because the latter influence timing but do not change relative probabilities of geometric trajectories. 
\vspace{-3mm}
\begin{figure}[!htb]
\scalebox{0.65}[0.65]{\includegraphics{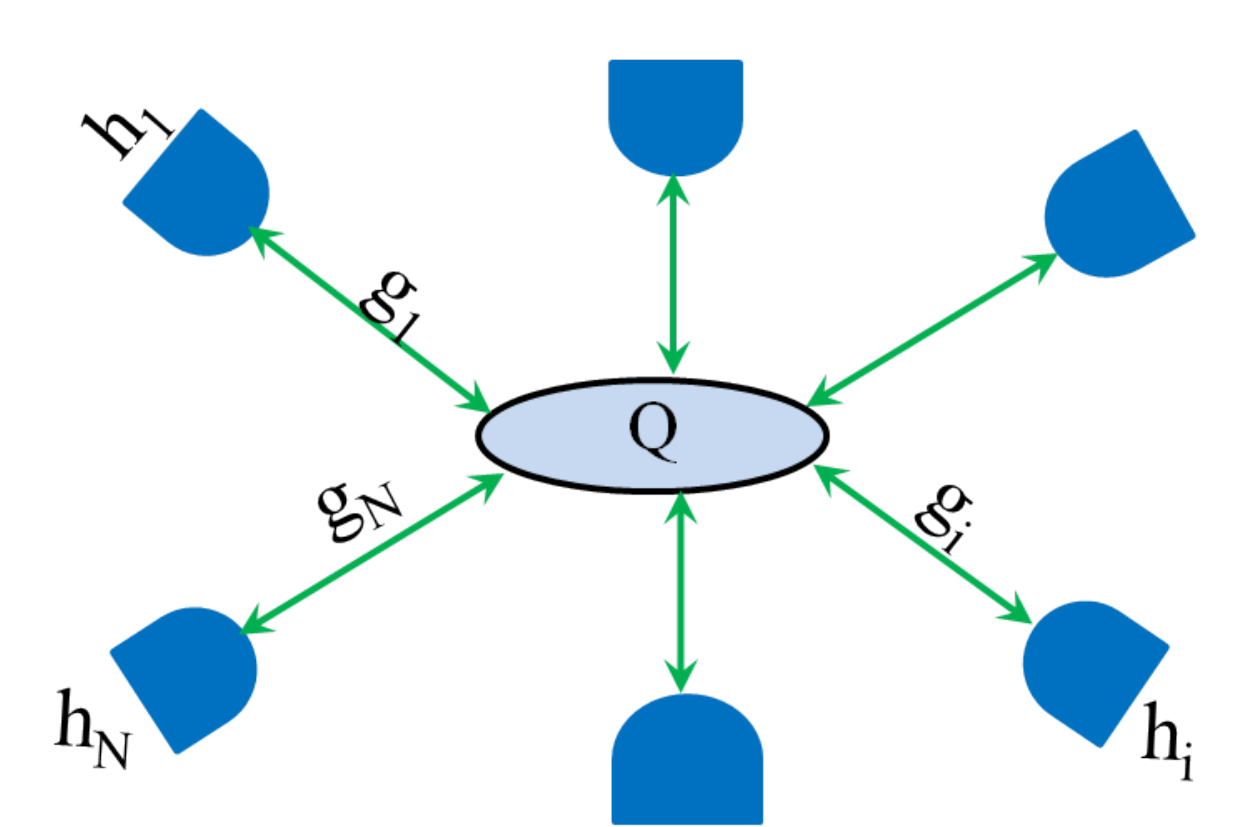}} \vspace{-0mm}
\hspace{1mm}   \caption{Chaotic cavity coupled  to $N$ electron reservoirs at different  potentials and temperatures.}
 \label{Cavity_main0}
\end{figure}

The structure of our article is as follows. In Section 2,
we will illustrate FRCCs in the  model of a chaotic cavity, shown in Fig.~\ref{Cavity_main0}, which frequently appears in studies of counting statistics \cite{altland-10prl}. In Section 3, we will  increase the complexity of the electric circuit geometry to demonstrate the ubiquity of FRCCs in mesoscopic electronics. Sections 4 and 5 play a supplementary role. They explore both presence and absence of FRCCs in specific models, which statistics of currents can be studied in  detail. They demonstrate that exactly solvable models produce results in agreement with more general theory developed in Sections 2 and 3.

\section{The model of chaotic cavity coupled to $N$ lead contacts} 
The model assumes that $N$ large  reservoirs  at different potentials  exchange electrons with a mesoscopic conducting region (cavity), in which the electron motion is randomized and is influenced by exclusion interactions due to the Pauli principle and Coulomb interactions \cite{altland-10prl}. 
The model also assumes that time-scales for self-averaging are much faster than the time-scale at which number of electrons in a cavity changes with time so that we can treat interactions with a mean field approximation. Electrons enter the cavity through the lead contacts with kinetic rates, $k_i^{\rm in}(Q) = h_i g_i f_1(Q)$, and leave the cavity  with rates  $k_i^{\rm out}(Q) = g_i f_2(Q)$, where $i=1,\ldots, N$. The parameter $h_i$
depends on thermal equilibrium conditions in the $i$-th reservoir,  $g_i$ is a strength of the cavity coupling to the $i$-th contact, and $Q$ is the instantaneous number of electrons in the cavity. The functions $f_1(Q)$ and $f_2(Q)$ describe the effect of many-body interactions on kinetic rates. 

We assume that all parameters, as well as functions $f_1$ and $f_2$, may depend on constant temperatures and potentials at reservoirs. 
As an example, first consider that the cavity is small (the quantum dot limit), so that it has only a single electron level at energy $E$. Coulomb interactions forbid to have more than one electron inside. The kinetic rate of escape of electron from this cavity into the $i$th contact can be estimated by the golden rule: $k_i^{\rm out}=2\pi Q \rho_i(E) |T_i(E)|^2 (1-1/[1+e^{(E-V_i)/(k_{\rm B}T)}])$, where $Q=1$ or $Q=0$ depending on the presence or absence of an electron inside the quantum dot. $T_i(E)$ is the element of the scattering matrix between the state inside the quantum dot and a state in the reservoir at the same energy $E$, $\rho_i(E)$ is the density of energy levels inside the $i$th contact near energy $E$, and the factor $1-1/[1+e^{(E-V_i)/(k_{\rm B}T)}]$ is due to the Pauli principle that forbids transitions into the filled states of the reservoir. Similarly, the kinetic rate of transitions from the $i$th reservoir into the quantum dot can be estimated as  $k_i^{\rm out}=2\pi (1-Q) \rho_i(E) |T_i(E)|^2 (1/[1+e^{(E-V_i)/(k_{\rm B}T)}])$. We can then identify 
$g_i=2\pi \rho_i(E) |T_i(E)|^2 (1-1/[1+e^{(E-V_i)/(k_{\rm B}T)}])$, $h_i=e^{-(E-V_i)/(k_{\rm B}T)}$, $f_1=(1-Q)$, and $f_2=Q$.
Another limit of the cavity model corresponds to a large cavity with kinetic rates induced by thermal over-barrier transitions and number of states inside the cavity much larger than the number of electrons. In such a classical limit, kinetic rates are given by Arrhenius form 
$k_i^{\rm in} \sim e^{(V_i-W_i)/k_{\rm B}T}$, and 
$k_i^{\rm out} \sim e^{(\mu(Q) -W_i)/k_{\rm B}T}$,
 where $W_i$ is size of the barrier that separates the cavity from the $i$th reservoir and $\mu(Q)$ is the chemical potential of electrons inside the cavity. Then we can identify $g_i=e^{ -W_i/k_{\rm B}T}$, $h_i = e^{V_i/k_{\rm B}T}$, $f_1=1$ and $f_2=e^{\mu(Q)/k_{\rm B}T}$. Note that, in both cases, 
at constant temperature,  parameters $h_i$ satisfy the relation 
\begin{equation}
h_i/h_j={\rm exp}\{ (V_i-V_j)/k_{\rm B}T \},
\label{hh}
\end{equation}
which guarantees the presence of the standard fluctuation relation (\ref{ft-2}).
This restriction, however, will not play any role in our following discussion.  

Let
\begin{equation}
Z({\bm \chi}) \equiv e^{S({\bm \chi})}=\sum_{\bm q} P( {\bm q} )   e^{ {\bm q} \cdot {\bm \chi} }
\label{gen-fun}
\end{equation}
be the generating function of currents through all leads. Here components of the $N$-vector, ${\bm q}$, are number of electrons that pass through corresponding contacts during the observation time. In the rest of the article, all introduced vectors will correspond to sets of elements indexed by the indices of the corresponding reservoirs, e.g. ${\bm \chi}=(\chi_1,\ldots \chi_N)$ is the vector of {\it counting parameters}. 
 $S({\bm \chi})$ is called the {\it cumulant generating function} because its knowledge corresponds to knowledge of all cumulants of the current distribution, e.g. 
\begin{equation}
\langle q_i \rangle = (\partial S/\partial \chi_i)_{{\bm \chi =0}}, \,\,\,  {\rm var}(q_i)= (\partial^2 S/\partial \chi_i^2)_{{\bm \chi =0}},\,\,\, {\rm etc}.
\label{cgf1}
\end{equation}
 We will detect FRCCs by setting all $\chi_i$, where $i\ne k $ for some $k \in 1,\ldots,N$, to zero and observing the symmetry,
\begin{equation}
S(0,..., 0,\chi_k,0,..., 0) = S(0,..., 0, -\chi_k+F_k,0,..., 0),
\label{ft-3}
\end{equation}
where $F_k$, for a given $k$, is a constant parameter. By applying the inverse Legendre transform, one can verify that (\ref{ft-3}) leads to the FRCC,
\begin{equation}
P(q_k)/P(-q_k) = {\rm exp} (q_kF_k).
\label{frcc}
\end{equation}

Let us calculate the probability, $p_{ij}$, of that the nearest entering electron will have a geometric trajectory that enters the cavity via the contact $i$ and leaves the cavity via the contact $j$. Kinetic rates for this particle depend on $Q$, which may change with time arbitrarily, however, the {\it ratio} of either two in-going or two outgoing kinetic rates for the given particle remains constant, which means that the probability of the geometric trajectory for this particle can be found explicitly:
\begin{equation}
p_{ij}=\frac{h_ig_i}{\sum_{k=1}^N h_kg_k } \times \frac{g_j}{\sum_{r=1}^N g_r }.
\label{pij}
\end{equation}
Let $P(n)$ be the probability that during a large observation time exactly $n$ electrons enter the cavity via any node. 
The probability of a geometric trajectory in (\ref{pij}) is independent of $n$ and of other particle's trajectories so the generating function $Z({\bm \chi})$ is given by
\begin{equation}
Z({\bm \chi}) = \sum_{n=0}^{\infty} P(n) \left( \sum_{i,j=1}^N p_{ij}e^{\chi_i-\chi_j} \right)^n. 
\label{zchi1}
\end{equation}
Although it is impossible to derive an explicit expression for $P(n)$, one can check that the symmetry (\ref{ft-3}) is the symmetry of each term in (\ref{zchi1}) with
\begin{equation}
F_k={\rm ln}\left(  \frac{    \sum_{i \ne k}^N g_ih_i    } { h_k \sum_{i \ne k}^N g_i } \right) , \quad k=1,\ldots,N,
\label{fk}
\end{equation}
which proves the FRCC for the cavity model.

Note that parameters ${\bm F}$ depend not only on lead characteristics ${\bm h}$, which  can be externally controlled, but also on coupling parameters ${\bm g}$. The latter may nontrivially depend on voltages due to appearance of screening charges in the vicinity of components of the nanoscale circuit at nonequilibrium conditions \cite{brouwer-01prb}. Estimation of parameters ${\bm g}$ from knowledge of lowest current cumulants may be difficult because functions 
$f_1(Q)$ and $f_2(Q)$ influence lowest current cumulants.
  Surprisingly, the FRCCs  do not depend on interactions encoded in functions $f_1(Q)$ and $f_2(Q)$ at all.   Hence measurements of FRCCs can provide us with a unique approach to measure the vector  ${\bm g}$ in a nonequilibrium regime irrespective of many-body interactions inside the cavity.

We also note that our derivation of the generating function is valid only for very large observation time, so that we could disregard the trajectories of electrons that entered but did not leave the cavity. Thus, FRCCs must be understood only in the sense of the dominating exponent (also known as the Large Deviation Function) of the probability distribution of a current.

\section{ Stochastic transport on networks}
To explore general principles that lead to FRCCs, we consider a generalization of the cavity model to a network of chaotic cavities coupled to  lead contacts and to each other, e.g. as shown in Fig.~\ref{back_main}.  Electrons enter cavities  through the reservoirs (with the rate $h_ig_i$ for the $i$-th cavity).  If two cavities (graph nodes) $i$ and $j$ are connected, each electron in the node $i$ can make a transition to node $j$ with rate $g_{ij} $; generally, $g_{ij}\ne g_{ji}$. Eventually, electrons leave the network through one of the contacts.   Physics of incoherent effects due to the shot noise and thermal Johnson-Nyquist noise in continuous conductors can be obtained from the 
 continuous limit of such network models \cite{cont-limit}.

For simplicity, we assume that all electrons are non-interacting, i.e. $f_1^i=1$ and $f_2^i=Q_i$ for all cavities, although generalizations to local interactions ($f_2^i(Q_i) \ne Q_i$) are possible because such interactions do not change relative probabilities of geometric trajectories \cite{GUC}. It was shown in \cite{GUC} that finding statistics of currents in such a model reduces to solving a finite set of coupled linear algebraic equations. Although explicit solutions are bulky and not illuminating, we  used them to check the presence/absence of an FRCC for any link of a network numerically, as we explain in Section 5.  As expected, we {\it did not} find an FRCC when there was no previously mentioned topological constraints on geometric trajectories. However, FRCCs were obtained in two wide classes of links with such constraints.

\subsection {Links that do not belong to any loop of the network}
These links represent lead contacts and also internal links of the network that on removal break the network into disjoint components. We marked such links by green color in Fig.~\ref{back_main}.
 Suppose that we can trace the geometric trajectory of a single electron. When electron enters the network it makes a single transition through a lead contact in positive direction and then makes one transition through one of the leads in negative direction when it leaves the network. Let $p_{kj}$ be the probability that the electron that enters through the contact $k$ leaves the network through the contact $j$.
If an electron enters through the contact $k$, then the moment generating function of currents through the contact $k$ that this electron produces during its life-time is given by
\begin{equation}
Z_k(\chi_k) = p_{kk} + \sum_{j\ne k} p_{kj}e^{\chi_k}.
\label{z1}
\end{equation}
The generating function of currents in the contact $k$, which is produced by an electron that enters from contact $j$, $j\ne k$, is given by
\begin{equation}  
Z_j(\chi_k) = \sum_{j'\ne k} p_{jj'} +  p_{jk}e^{-\chi_k}.
\label{z2}
\end{equation}

\vspace{-0.3cm}
\begin{figure}[!htb]
\scalebox{0.55}[0.55]{\includegraphics{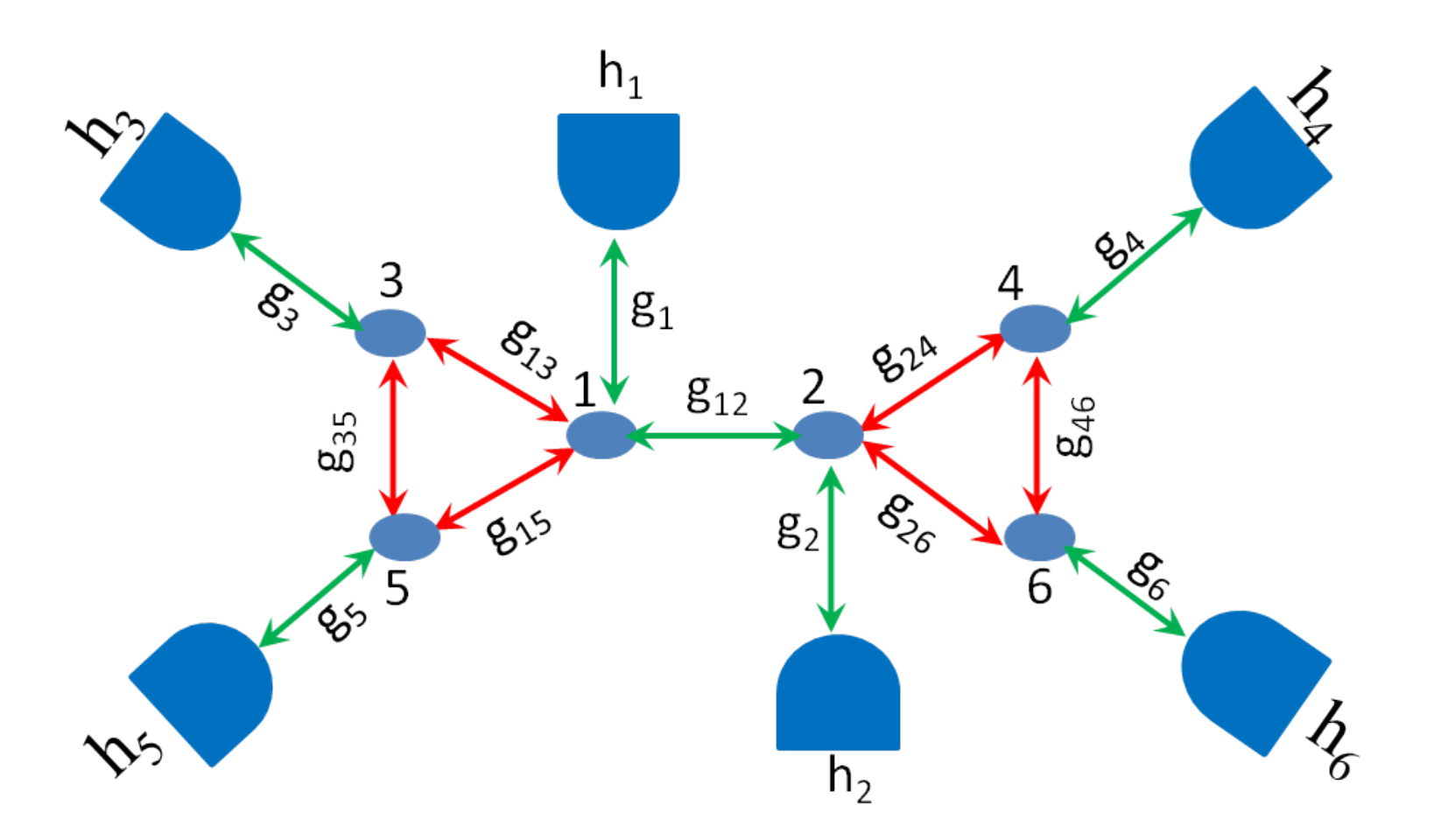}} \vspace{0.3mm}
\hspace{1mm}   \caption{Circuit of coupled chaotic cavities and electron reservoirs. Green color marks links that carry currents that satisfy FRCCs. Red color marks links with currents that violate FRCCs.} \label{back_main}
\end{figure}
  
 Let $P_{q_1,\ldots,q_N}(t)$ be the probability that during time $t$ exactly $q_1,\ldots,q_N$ particles enter the network through the contacts $1,\ldots, N$. Since particles enter independently,  this distribution is, in fact, Poisson and its generating function is given by
\begin{equation} 
Z_P(\bm s)\equiv \sum_{{\bm q}} P_{{\bm q}}e^{{\bm s}{\bm q}}={\rm exp}\left( \sum_{i=1}^N h_ig_i t (e^{s_i}-1)  \right).
\label{z3}
\end{equation}
 Since electrons do not interact,  probabilities $P_{q_1,\ldots,q_N}(t)$  and $p_{kj}$ are not correlated.  The statistics of currents through contact $k$ during the whole process is then given by 
\begin{equation}
Z(\chi_k)=\sum_{\bm q}P_{q_1,\ldots,q_N} Z_1^{q_1}\cdot \ldots \cdot Z_N^{q_N},
\label{z4}
\end{equation}
 where we used the fact that the generating function of a sum of independent processes is the product of generating functions of individual processes.  The latter expression shows that $Z(\chi_k)$ coincides with $Z_P({\bm s})$ up to a change of variables $e^{s_i} \rightarrow Z_i(\chi_k)$ for any $i$.
 This fact and the explicit expressions for  $Z_P({\bm s})$ and $Z_i(\chi_k)$ lead us to the conclusion that the total generating function of currents depends on $\chi_k$ only through the combination, $ h_kg_k \sum_{j\ne k} p_{kj}e^{\chi_k} + \sum_{j\ne k} h_j g_j  p_{jk}e^{-\chi_k}$, which has the symmetry under the change of variables, $\chi_k \rightarrow -\chi_k + F_k$, where
 \begin{equation}
 F_k={\rm ln} \left( \frac{ \sum_{j\ne k}  h_j g_j  p_{jk} } {h_kg_k \sum_{j\ne k} p_{kj}}  \right).
 \label{fk2}
 \end{equation}
The case with a link that connects two otherwise disjoint components, $\alpha$ and $\gamma$, is proved similarly: Let $p_{\alpha}$ be the probability that  a given particle that enters the network in the  components $\alpha$ leaves the network through some contact in another graph component $\gamma$. The generating function of currents through the link that connects such components is $Z_{\alpha}(\chi)=(1-p_{\alpha})+p_{\alpha} e^{ \chi}$. The rest of the proof is the same as for the currents through lead contacts, where instead of contact indices $k$ and $j$ we  write component indices $\alpha$ and $\gamma$.

\subsection{Quantum coherence among trajectories}
So far we assumed the lack of quantum interference among particle trajectories. The arguments leading to (\ref{ft-3}) and (\ref{fk2}) for a particle's motion on a network, however, do not refer to any classical or thermodynamic reason. For example, one can  imagine that the distribution of electrons in leads is not at equilibrium and that probabilities $p_{ij}$ are influenced by quantum interference of different trajectories that connect nodes $i$ and $j$.
We just should assume that (a) events of particle's escapes into reservoirs destroy coherence, (b) particles enter the network according to the Poisson statistics, and (c)  transition probabilities $p_{ij}$ are constant. Such conditions are realized in the quantum regime when the single electron scattering amplitude, $s_{ij}(E)$, between any pair of different reservoirs $i$ and $j$ in a channel with any energy $E$, is small, i.e. $|s_{ij}(E)| \ll 1$. In this limit, we can disregard simultaneous multi-electron scattering processes, although purely quantum effects, such as quantum interference of trajectories, can still influence $s_{ij}$.

The generating function for $N$ terminals and non-interacting fermions was derived by Levitov and Lesovik \cite{levitov}. In their determinant formula for generating function, we can set only a single counting parameter, $\chi_k$, to be non-zero. Then, for arbitrary scattering matrix, the cumulant generating function of a current through a corresponding reservoir is given by 
\begin{equation}
S(\chi_k)={\rm ln} \left (\prod_E(C_k(E)+A_{k}(E)e^{\chi_k}+B_{k}(E)e^{-\chi_k}) \right),
\label{z5}
\end{equation}
 where $A_k$, $B_k$, and $C_k$ are constants that depend on multielectron scattering amplitudes and channel populations.
According to the golden rule, in the limit of weak transmission, coefficients $A_k$ and $B_k$ can be approximated to first order in $| s_{jk}|^2$ as 
\begin{equation}
A_k(E)=\sum_{j\ne k}n_k(E)(1-n_j(E))|s_{kj}|^2
\label{z7}
\end{equation}
and 
\begin{equation}
B_k(E)=\sum_{j\ne k}(1-n_k(E))n_j(E)|s_{jk}|^2.
\label{z8}
\end{equation}
Here $n_i(E)$ is population of the channel with energy $E$ in the $i$-th contact.
 Moreover, in this limit, we can use ${\rm ln}(1+x) \approx x$ to approximate   
\begin{eqnarray}
\nonumber S(\chi_k) \approx \sum_{j\ne k} \sum_E |s_{kj}|^2(1-n_k)n_j (e^{-\chi_k}-1)+\\
\sum_{j\ne k}\sum_E |s_{kj}|^2n_k(1-n_j)(e^{\chi_k}-1).
\label{z16}
\end{eqnarray}
 Hence, the counting statistics of individual contact currents of non-interacting electrons is equivalent to a sum of two Poisson processes, 
\begin{equation}
S(\chi_k)=A_k(e^{\chi_k}-1)+B_k(e^{-\chi_k}-1).
\label{z6}
\end{equation}
 This generating function is symmetric under exchange $\chi_k \rightarrow -\chi_k +F_k$ and leads to an FRCC with
\begin{equation}
F_k={\rm ln} \left( \frac {\sum_E(1-n_k(E))\sum_{j\ne k}|s_{kj}(E)|^2n_j(E)  }{    \sum_En_k(E)\sum_{j \ne k}|s_{kj}(E)|^2(1-n_j(E))} \right).
\end{equation}


\subsection {Networks connected to a single reservoir}
A circuit that is coupled to a single particle reservoir, such as shown  in Fig.~\ref{Cyclic_main}(a,b), is another class of systems that show FRCCs for currents through some links that, in contrast to previous case, belong to loops of the graph. Such networks can be used to describe statistics of single molecule events \cite{astumian-11rev}. For example, the graph in Fig.~\ref{Cyclic_main}(a) corresponds to kinetics of a biological enzyme \cite{astumian-11rev} where the external link corresponds to the process of enzyme creation/degradation \cite{GUC}. 
 Probabilities of geometric trajectories in such open networks depend only on  kinetic rates and not on time moments of individual transitions through links \cite{GUC}.
\begin{figure}[!htb]
\scalebox{0.55}[0.55]{\includegraphics{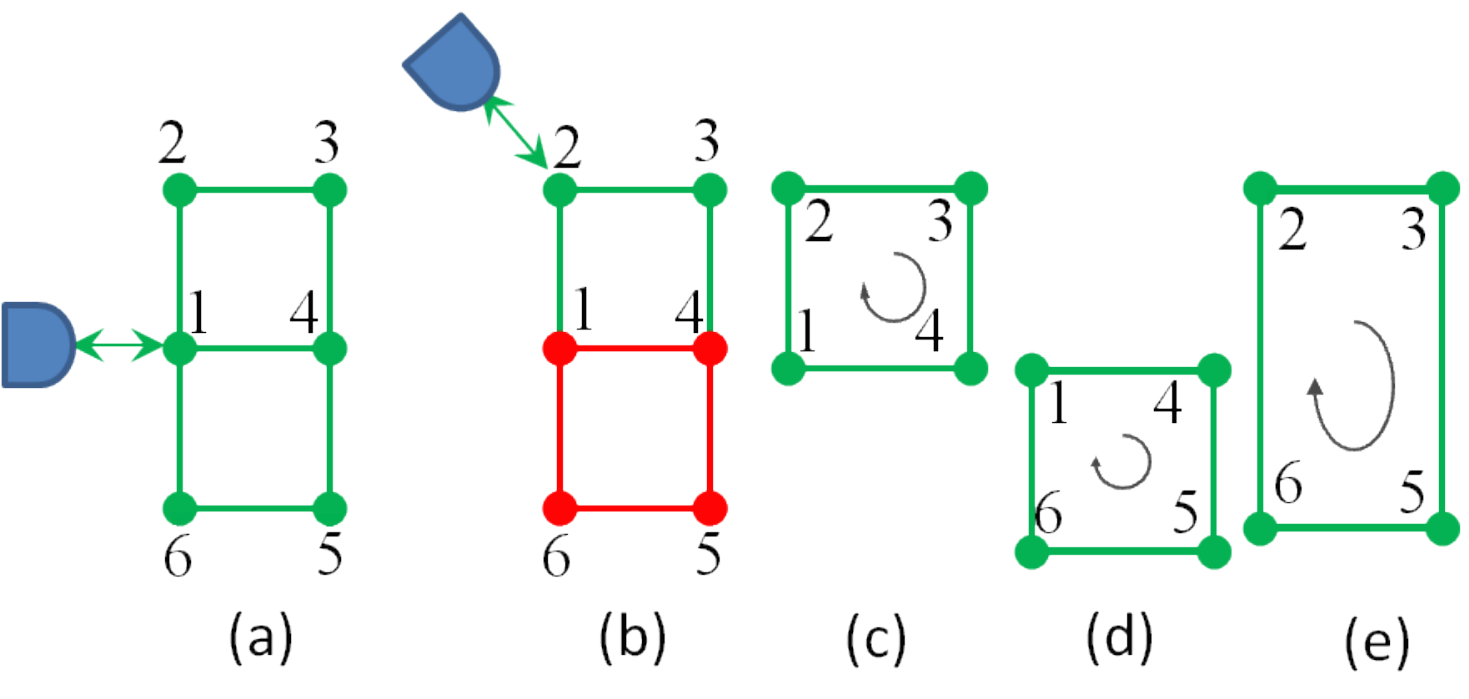}} \vspace{-3mm}
\hspace{1mm}   \caption{(a) and (b): Networks coupled to a single reservoir. Green links  carry currents that satisfy FRCC and red links carry currents that generally do not satisfy FRCC. (c), (d) and (e): Distinct cycles of the graph. Cyclic arrows define "+" directions of cycles.} \label{Cyclic_main}
\end{figure}
Consider, for example, the model in Fig.~\ref{Cyclic_main}(a) where we will be interested in currents through the link that connects nodes $4$ and $1$. The arguments in Section 3A lead for this model to a conclusion that it is sufficient to prove the FRCC for currents produced by a single particle during its life-time because the symmetry of a single particle generating function becomes the symmetry of the full counting statistics of currents when there is only one external particle reservoir. Consider a particle that is just appeared on the node 1 on a graph in Fig.~\ref{Cyclic_main}(a). Let $r$ be the kinetic rate of the transition of this particle from node 1 to the reservoir and let $k_{ij}$ be kinetic rates of transitions between a pair of nodes, $i$ and $j$. From the node 1, the particle can return to the reservoir with probability 
\begin{equation}
p_{{\rm esc}}=r/(r+k_{12}+k_{16}+k_{14}),
\label{z10}
\end{equation}
 producing no currents in the system, or it can move into another node. In the latter case, we know that the particle must eventually return to the node 1 because it is the only place from which it can escape from the network, and our measurement time is assumed to be much larger than the particle's life-time on the graph.

Each particle's return to node 1 corresponds either to making none or one of three possible cycles \cite{astumian-11rev},  shown in Fig.~\ref{Cyclic_main}(c,d,e). Each cycle can be passed in two directions, that we will mark "$+$" or "$-$".
Let $p_0$, $\eta_{1\pm}$, $\eta_{2\pm}$, and $\eta_{3\pm}$ be probabilities of such return events, where indices $1,2,3$ correspond to cycles in Fig.~\ref{Cyclic_main}, respectively, (c), (d), and (e). These probabilities depend only on kinetic rates, i.e. they are constants \cite{GUC}.
The generating function of currents through the link $(4,1)$, produced by all geometric trajectories that make exactly one return to node 1 is then given by
\begin{eqnarray}
\nonumber Z_1(\chi_{41})=(1-p_{{\rm esc}})p_{{\rm esc}} [p_0+\eta_3+\eta_{-3}+\\
(\eta_1+\eta_{-2})e^{\chi_{41}}+(\eta_{-1}+\eta_{2})e^{-\chi_{41}}].
\label{zz14}
\end{eqnarray}
 After each return to the node 1 the process renews. Hence the generating function produced by all trajectories that make exactly $n$ returns to the node 1 is  given by
\begin{eqnarray}
\nonumber Z_n(\chi_{41})=p_{{\rm esc}}[ (1-p_{{\rm esc}}) (p_0+\eta_3+\eta_{-3}+\\
(\eta_1+\eta_{-2})e^{\chi_{41}}+(\eta_{-1}+\eta_{2})e^{-\chi_{41}})]^n.
\label{zz15}
\end{eqnarray}
 The total single particle generating function is the sum over current generating functions, induced by all geometric trajectories, i.e. 
\begin{equation}
Z(\chi_{41})=\sum_{n=0}^{\infty}Z_n(\chi_{41}),
\label{z17}
\end{equation}
 which explicitly can be written in a form
\begin{equation}
Z(\chi_{41})=p_{{\rm esc}}/[1-(1-p_{{\rm esc}})(Ae^{\chi_{41}}+Be^{-\chi_{41}}+C)],
\label{zchi}
\end{equation}
where $A=\eta_1+\eta_{-2}$, $B=\eta_{-1}+\eta_{2}$, $C=p_0+\eta_3+\eta_{-3}$. Eq. (\ref{zchi}) is invariant under change of variables, $\chi_{41} \rightarrow -\chi_{41} +F_{41}$, where 
\begin{equation}
F_{41}={\rm ln}([\eta_{-1}+\eta_{2}]/[\eta_1+\eta_{-2}]),
\label{z9}
\end{equation}
 which completes our derivation of the FRCC for the link $(4,1)$.  One can easily extend our arguments to all green links in Figs.~\ref{Cyclic_main}(a,b). 

\section{Exactly solvable models}

In this section, we derive explicit expression for the generating function of currents in the cavity model in several important limits and check the presence of FRCC explicitly. We show that results are in full agreement with the symmetry described by (\ref{ft-3}) and (\ref{fk}).   

\subsection{Stochastic path integral solution of chaotic cavity model}
To derive FRCCs for the model in Fig.~\ref{Cavity_main0}, we will employ the stochastic path integral technique \cite{altland-10prl,sinitsyn-09pnas,sinitsyn-07prl} that was previously applied to calculations of current cumulants and studies of standard fluctuation theorems in the cavity model.  This approach is applied to the case of the cavity  that has a mesoscopic size so that typically we have $Q \gg 1$.  Following this  approach, the counting statistics at a steady state is
$S({\bm \chi}) = Ht$, where $H=H({\bm \chi},Q_c({\bm \chi}),\chi_c({\bm \chi}))$ is given by
\begin{equation}
H = \sum_{i=1}^N k_{i}^{\rm in}(Q_c) (e^{\chi_i+\chi_c}-1) +
k_{i}^{\rm out}(Q_c) (e^{-\chi_i-\chi_c}-1),
\label{H}
\end{equation}
where $Q_c$ and $\chi_c$ are expressed through ${\bm \chi}$ by solving steady state "Hamiltonian equations", 
\begin{equation}
\partial H/\partial Q_c=0, \quad \partial H/\partial \chi_c = 0.
\label{z12}
\end{equation}
 To explore those equations, it is convenient to introduce combinations of parameters,
\begin{equation}
a({\bm \chi}) = \sum_{i=1}^N h_i g_i {\rm exp} (\chi_i),
\label{aa4}
\end{equation}
\begin{equation} 
b({\bm \chi}) = \sum_{i=1}^N g_i {\rm exp} (-\chi_i).
\label{b4b}
\end{equation}
 Hamiltonian equations then explicitly lead to relations:
\begin{equation}
{\rm exp }(\chi_c) = [bf_2/a f_1]^{1/2},
\label{aa41}
\end{equation}
\begin{equation}
f_1'(\sqrt{ab}\sqrt{f_2/f_1}-a({\bm 0}))+f_2' (\sqrt{ab}\sqrt{f_1/f_2}-b({\bm 0})) =0,
\label{bb41}
\end{equation}
where functions $f_1$ and $f_2$ were defined in Section 2. 
 Generally, such nonlinear equations cannot be solved explicitly to determine $Q_c$ and $\chi_c$, but they do imply that $Q_c$, as well as combinations $ae^{\chi_c}$ and $b e^{-\chi_c}$, and hence $H$ and $S$ depend on counting parameters only via the product, $a({\bm \chi}) b({\bm \chi})$. It is then straightforward  to verify that  the symmetry, $S(0,\ldots,\chi_k,\ldots, 0)=S(0,\ldots,-\chi_k+F_k,0,\ldots)$, is also the symmetry of $a({\bm \chi}) b({\bm \chi})$ with
\begin{equation}
F_k={\rm ln}\left(  \frac{    \sum_{i \ne k}^N g_ih_i    } { h_k \sum_{i \ne k}^N g_i } \right) , \quad k=1,\ldots,N,
\label{fk4}
\end{equation}
which proves the FRCC for the cavity model in the limit of mesoscopic cavity size.

The scope of the path integral technique is limited to mesoscopic systems. However, we could explicitly verify that the FRCC  holds true in two exactly solvable limits. In the following subsections, we show exactly that Eq.  (\ref{fk}) is satisfied for a cavity with exclusion interactions ($f_1=Q_{\rm max}-Q$, $f_2=Q$), and in the classical limit  in which the number of available states in the cavity is much larger than the number of electrons so that in-going rates are not influenced by the Pauli principle ($f_1=1$  and arbitrary $f_2(Q)$). 

\subsection{Exact solution of cavity model with exclusion interactions}
Here we consider the cavity model with $N$ lead contacts. Electrons do not interact with each other except via the exclusion interactions due to the Pauli principle. Electrons enter the cavity through a lead contact $i$ ($i=1,\ldots, N$) with kinetic rate, $k_i^{\rm in}(Q_c) = h_i g_i (Q_{\rm max}-Q_c)$, and  leave the cavity  with rate $k_i^{\rm out}(Q_c) = g_i Q_c$. The latter rate is proportional to the number of electrons $Q_c$ in the cavity. This corresponds to non-interacting case. 
We are interested in statistics of currents through a specific lead contact, $k$.
Our exact solution of this model is based on the observation that the model is equivalent to the model of independent currents through $Q_{\rm max}$ quantum dots \cite{sinitsyn-07prl,ohkubo-09pre}. Each dot can have either zero or maximum one electron inside. If there is no electron in the dot then with rates $h_i g_i$, where $i=1,\ldots, N$, an electron jumps into the dot. If the dot has electron inside, then with rates  $g_i$, $i=1,\ldots, N$, it leaves to one of the leads. Calculation of the cumulant generating function at steady state in such a two-state model is straightforward and was discussed in number of publications, e.g. \cite{derrida-98prl,nazarov-03prb,szabo-06jcp,sinitsyn-07epl} with a minor difference that here we assume many lead contacts. Following e.g. \cite{szabo-06jcp}, the cumulant generating function is given by the larger eigenvalue of the matrix

\begin{equation}
H=
\left( \begin{array}{cc}
-\sum_{i=1}^N h_i g_i                                         & \left(\sum_{i \ne k}^N  g_i   \right)+ g_ke^{-\chi_k}\\
\\
\left( \sum_{i \ne k}^N h_i g_i \right)  + h_kg_ke^{\chi_k} & -\sum_{i=1}^N g_i
\end{array} \right).
\label{h}
\end{equation}
Its eigenvalue, $\lambda_0$, can be explicitly written as
\begin{equation}
\lambda_0=\frac{1}{2}(-K+\sqrt{K^2+4g_k(h_kg_k + e^{-\chi_k}A + e^{\chi_k} h_kB }),
\label{lam}
\end{equation}
where $A=\sum_{i \ne k}^N h_i g_i $, $B=\sum_{i \ne k}^N g_i $, and $K$ is independent of $\chi_k$ constant. Since the model of the cavity with exclusion interactions is equivalent to $Q_{\rm max}$
independent processes, each having counting statistics $\lambda_0(\chi)t$, the cumulant generating function for the complete model is given by

\begin{equation}
S(\chi_k)=Q_{\rm max} \lambda_0(\chi_k) t.
\label{cgf1}
\end{equation}
Obviously, $\lambda_0(\chi)$ is symmetric under exchange $\chi_k \rightarrow -\chi_k + F_k$, where $F_k$ is given by

\begin{equation}
F_k={\rm ln}\left(  \frac{    \sum_{i \ne k}^N g_ih_i    } { h_k \sum_{i \ne k}^N g_i } \right),
\label{fk-supl}
\end{equation}
in agreement with  (\ref{fk}).

\subsection{Exact solution of the cavity model for stochastic transitions with local interactions}
 Another exactly solvable model corresponds to a system of locally interacting particles performing stochastic transitions  through a cavity connected to $N$ reservoirs (See Fig. \ref{Cavity_main0}) with constant in-going kinetic rates.  Let a state vector, $|Q_{c}\rangle$, be determined by an occupation number $Q_{c}$, associated with the cavity. Kinetic rates for transitions from the cavity to $i$-th reservoir is given by 
\begin{equation}
k_{i}^{\rm out} = Q_c f_c(t,Q_c) g_{i},
\label{kk1}
\end{equation}
 where  $f_c(Q_c)$  describes arbitrary  local  repulsive interactions inside the cavity that influence out-going kinetic rates by renormalizing single electron free energy in the mean field approximation. The in-going kinetic rates through the $i$-th reservoir is
\begin{equation}
k_i^{\rm in}=h_ig_i.
\label{kk2}
\end{equation}
 Here, for generality, we can also allow arbitrarily prescribed explicit periodic time-dependence  of  ${\bm h}=(h_1,h_2,\ldots h_N)$ with period $\tau$, keeping other parameters constant. A particle distribution function $P({ Q_c})$ can be written as a state vector 
\begin{equation}
\Psi=\sum_{{Q_c}}P({ Q_c})|{Q_c}\rangle
\label{psi1}
\end{equation}
 that satisfies the master equation
\begin{equation}
\partial \Psi/\partial t=\hat{{\cal L}}\Psi.
\label{psi2}
\end{equation}
 To derive the counting statistics of currents, we should consider evolution with a {\it twisted master operator}, $\hat{{\cal L}}_{{\bm\chi}}$ that can be obtained from the operator $\hat{{\cal L}}$ by multiplying its off-diagonal elements by factors $e^{\pm \chi_i}$ to count transitions from/to reservoir $i$. For details see Refs.~\onlinecite{GUC,nazarov-03prb}. Following those rules, for our cavity model, we obtain the form of the twisted master operator:
\begin{eqnarray}
\label{master-operator-twisted} \hat{{\cal L}}_{{\bm\chi}}=-\sum_{j}g_{j}[\hat{a}_{c}^{\dagger}(f_c (\hat{Q}_c) \hat{a}_{c}-&h&_{j}e^{\chi_{j}})\\
+h_j-f_c(\hat{Q}_c) \hat{a}_c &e&^{-\chi_{j}}], \quad
\hat{Q}_c=\hat{a}_{c}^{\dagger}\hat{a}_{c},
\end{eqnarray}
where we have used the "second quantized"  version of the master equation with 
\begin{equation}
\hat{a}_{c}^{\dagger}|Q_{c}\rangle=|Q_{c}+1\rangle, \quad \hat{a}_{c}|Q_{c}\rangle=Q_{c}|Q_{c }-1\rangle
\label{kk3}
\end{equation}
 being the creation/annihilation operators. Here we note that operator $\hat{{\cal L}}_{{\bm\chi}}$ in (\ref{master-operator-twisted}) is generally non-quadratic in $\hat{a}_{c}^{\dagger}$ and $\hat{a}_{c}$, which is the result of many-body interactions inside the cavity. For the case with local particle interactions,  it becomes easier to obtain the solutions for the master equation by looking not at evolutions of ket-vector but rather at the backward in time evolution of bra-vector $\langle \Psi_{{\bm\chi}}(t)|$ given by \cite{GUC}
\begin{equation}
\frac{\partial \langle \Psi_{{\bm\chi}}\vert}{\partial t} =-\langle \Psi_{{\bm\chi}}\vert  \hat{{\cal L}}_{{\bm\chi}} (t),  \;\; \langle \Psi_{{\bm\chi}}(t+\tau)|=\langle \Psi_{{\bm\chi}}(t)|e^{-S(\tau)}.
\label{bra-evolution_sup}
\end{equation}
According to \cite{nazarov-03prb,szabo-06jcp,GUC}, the largest $S(\tau)$ in (\ref{bra-evolution_sup}) coincides with the cumulant generating function (CGF) of currents per period of the parameter evolution in the $t \to \infty$ limit. Following \cite{GUC}, we will search for the solution of (\ref{bra-evolution_sup}) in the form of a coherent state bra-vector 
\begin{equation}
\langle \Psi_{{\bm\chi}}(t)\vert = \langle {0}\vert \exp(\zeta_{c}\hat{a}_{c})e^{-S(t)},
\label{kk4}
\end{equation}
and substitute this ansatz in Eq. (\ref{bra-evolution_sup}).
Using the property that  
\begin{equation}
\langle \Psi_{{\bm\chi}}(t)\vert \hat{a}_c^{\dagger}=\langle \Psi_{{\bm\chi}}(t)\vert \zeta_{c}, 
\label{kk5}
\end{equation}
and then grouping separately terms near $\langle \Psi_{{\bm\chi}}(t)\vert f_c(\hat{Q}_c)\hat{a}_c$ and separately the remaining functions that multiply   $\langle \Psi_{{\bm\chi}}(t)\vert$, we find that Eq. (\ref{bra-evolution_sup}) is satisfied if
\begin{eqnarray}
\label{eigenvalue-bra-coherent} \zeta_{c}= \frac{\sum_{j}g_{j}e^{-\chi_{j}}}{ \sum_{j}g_{j}},
\end{eqnarray}
\vspace{-0.5cm}
\begin{eqnarray}
dS/dt=\sum_{j} g_{j} h_{j}(t)(e^{\chi_j}\zeta_{c}-1).
\label{bra-CGF}
\end{eqnarray}
Only parameters $h_j$ in (\ref{bra-CGF}) are time-dependent. It is then trivial to integrate (\ref{bra-CGF}) to find
\begin{eqnarray}
S(\tau,{\bm \chi }) = \sum_{j} g_{j}\bar{h}_{j}(e^{\chi_j}\zeta_{c}-1),
\label{bra-CGF1}
\end{eqnarray}
where we defined $\bar{h}_{j} \equiv \int^{\tau}_{0}h(t')dt'$.
We can now write $S(\tau,{\bm \chi })$ explicitly as
\begin{equation}
S(\tau,\chi_k)=\frac{\sum_{jk}\bar{h}_{k}g_{k}g_{j}(e^{-\chi_{j}+\chi_{k}}-1)}{\sum_{j}g_{j}}.
\end{equation}

To illustrate the presence of the FRCC, we write $S(\tau,{\bm \chi })$ for individual counting fields by setting $\chi_{j}=0$ for $j\ne k$, then
\begin{eqnarray}
S(\tau,\chi_k)=C+B_{k}e^{-\chi_{k}}+A_{k}e^{\chi_{k}},
\label{ftic_form_S}
\end{eqnarray}
where $C=\frac{\sum_{jk}\bar{h}_{j}g_{k}g_{j}}{\sum_{j}g_{j}}$, and
\begin{equation}
A_{k}=\frac{\bar{h}_{k}g_{k}\sum_{j \ne k}g_{j}}{\sum_{j}g_{j}}, \, B_{k}=\frac{g_{k}\sum_{j\ne k}\bar{h}_{j}g_{j}}{\sum_{j}g_{j}}.
\label{kk7}
\end{equation}
 The form of CGF in Eq.~(\ref{ftic_form_S}) has the symmetry under the exchange of $\chi_k \rightarrow -\chi_k + F_k$, where $F_k$ is given by
\begin{equation}
F_k={\rm ln}\left(  \frac{    \sum_{j \ne k}^N g_j\bar{h}_j    } { \bar{h}_k \sum_{j \ne k}^N g_j } \right).
\label{fj_cavityl}
\end{equation}
This expression for $F_k$ is the same as (\ref{fk}). 

\section{Numerical check of FRCC for networks}

In this section, we consider more complex networks and analyze the validity of the FRCC for any particular link numerically.
\subsection{Network with loops and backbone link}
\begin{figure}[!htb]
\scalebox{0.7}[0.7]{\includegraphics{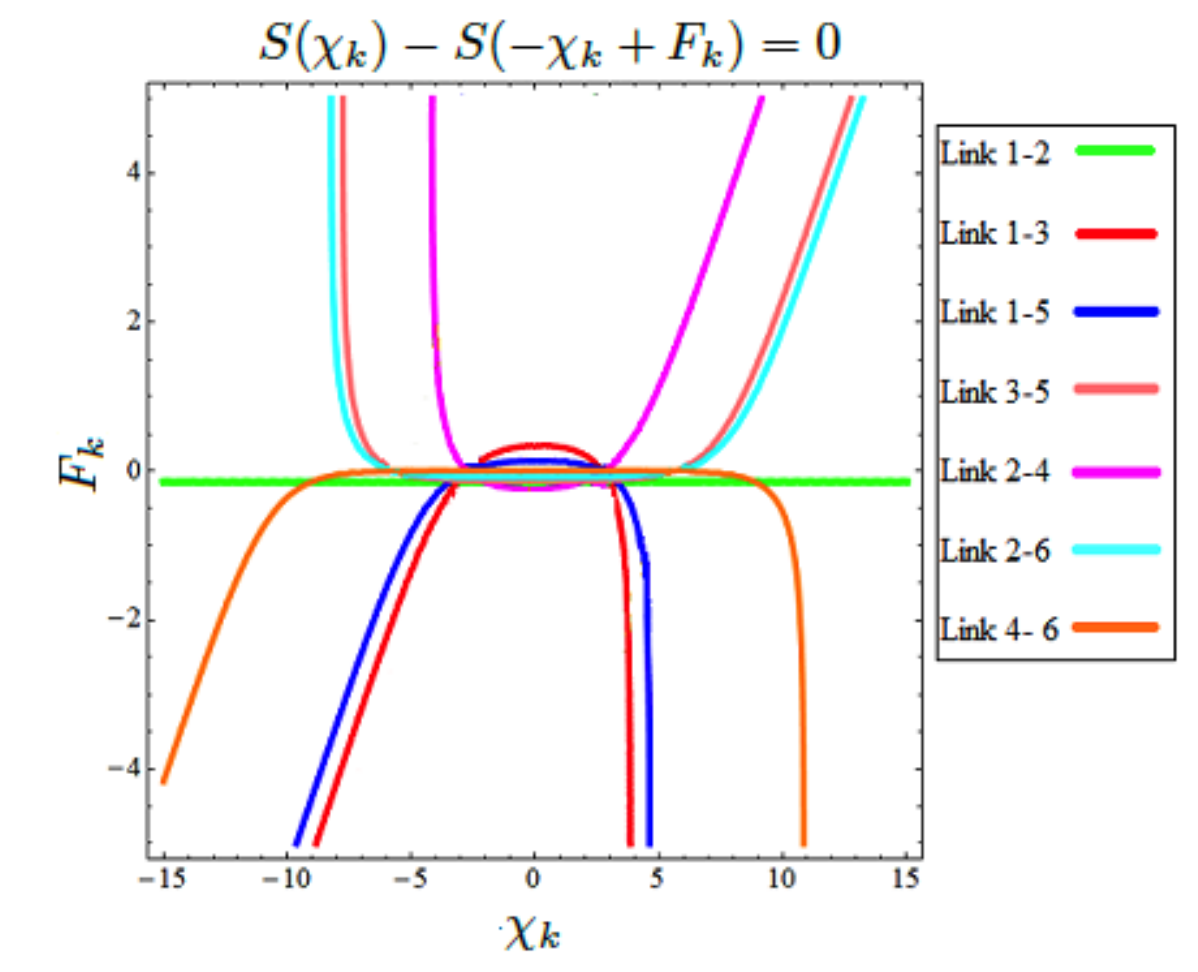}} \vspace{-3mm}
\hspace{1mm}   \caption{Numerical plots for the solution contours of $S(\chi)-S(-\chi+F)=0$ for currents in geometry of Fig.~\ref{back_main}. Parameters were set to numerical values: $g_{12}=0.3854$,  $g_{13}=0.6631$, $g_{15}=0.5112$, $g_{35}=0.6253$, $g_{24}=0.17483$, $g_{26}=0.02597$, $g_{46}=0.931946$, $g_{ij}=g_{ji}$, $g_{1}=0.5379$,  $g_{2}=0.004217$, $g_{3}=0.8144$, $g_{4}=0.8396$, $g_{5}=0.1765$, $g_{6}=0.8137$, $h_{1}=8.7139$,  $h_{2}=8.8668$, $h_{3}=4.6144$, $h_{4}=8.1632$, $h_{5}=7.6564$, and $h_{6}=8.01407$.} \label{6noderesults}
\end{figure}
Consider a graph that is coupled to external particle reservoirs and has loops and a backbone link as shown in Fig.~\ref{back_main}. For numerical check we restrict to models with only one reservoir per node. Generalizing the cavity model we define additional parameters that are the kinetic rates of transitions from node $i$ to node $j$, given by $k_{ij}=Q_i f_i g_{ij}$.
The twisted master operator for this case can be written as \cite{GUC}
\begin{eqnarray}
\label{master-operator-twisted_backbone} \hat{{\cal L}}_{{\bm\chi}}&=&\sum_{jk}g_{jk}\hat{a}_{j}^{\dagger}(e^{\chi_{jk}}f_k \hat{a}_{k}
-f_j\hat{a}_{j}) \nonumber \\
&-&\sum_{j}g_{j}[\hat{a}_{j}^{\dagger}(f_j \hat{a}_{j}-h_{j}e^{\chi_{j}})+h_j-f_j \hat{a}_je^{-\chi_{j}}],
\end{eqnarray}
where we introduced additional parameters $\chi_{ij}$ to count currents through internal links $(i,j)$.
   Following the prescription for the case of the single cavity (node) model we obtain the CGF, $S(\chi_{jk},\chi_{k})$, which reads:
\begin{eqnarray}
S(\tau,{\bm \chi }) =\sum_{j} g_{j}\bar{h}_{j}(e^{\chi_j}\zeta_{j}-1),
\label{bra-CGF_backbone}
\end{eqnarray}
where parameters $\zeta_{j}'s$ can be obtained by the solving following set of linear equations:
\begin{eqnarray}
\label{eigenvalue-bra-coherent_backbone} \sum_{j}g_{jk}(e^{\chi_{jk}}\zeta_{j}-\zeta_{k})-g_{k}(\zeta_{k}-e^{-\chi_{k}})=0.
\end{eqnarray}

Explicit functional form of $S$, even for a circuit of a moderate size in Fig.~\ref{back_main},  is quite bulky to be shown here. In any case, it is difficult to observe the presence of an  FRCC just by looking at an analytical expression for a generating function. To demonstrate the presence of an FRCC, we resort to numerical  solution of (\ref{bra-CGF_backbone}), (\ref{eigenvalue-bra-coherent_backbone}). We set all but one counting parameters in  (\ref{bra-CGF_backbone}) and (\ref{eigenvalue-bra-coherent_backbone}) to be zero. The presence of an FRCC can be checked by looking at the solution contours of the equation $S(\chi)-S(-\chi+F)=0$ plotted as a function, $F(\chi)$. An FRCC occurs for the link $k$ if there exists a  solution contour for which $F_k$ is independent of $\chi_{k}$. In other words, if we plot all pairs, ($\chi_k,F_k$), that satisfy equation,  $S(\chi_k)-S(-\chi_k+F_k)=0$, in plane with axes $\chi_k$ and $F_k$, then if an FRCC holds, the curve must be a horizontal line, which is parallel to $\chi_k$-axis. 

The plot in Fig.~\ref{6noderesults} shows a solution $F_{ij}(\chi_{ij})$, where $(i,j)$ runs through all internal links of the network in Fig. \ref{back_main}.
It clearly shows  that only for the link $(1,2)$ we have $F_{12}={\rm const}$.  For all other links, solution contours are nontrivial functions ($F_{ij}(\chi_{ij}) \ne {\rm const}$).   The FRCC is upheld by the links  that do not belong to any loop. Such a link in our example is the $(1 ,2)$ link, and this is the only link that supports FRCC, which is in agreement with our discussion in the main text. 

\subsection{Numerical check for networks coupled to a single reservoir}
In this subsection, we numerically analyze the class of networks with loops. Particles can enter and leave only via a single reservoir.  Figs.~\ref{Cyclic_main}(a,b) show networks, in which all internal links belong to some loops of a graph. Detailed balance on kinetic rates is not assumed. The twisted master operator for this case is 
\begin{eqnarray}
\label{master-operator-twisted_cycle} \hat{{\cal L}}_{{\bm\chi}}&=&\sum_{jk}g_{jk}\hat{a}_{j}^{\dagger}(e^{\chi_{jk}}f_k \hat{a}_{k}
-f_j\hat{a}_{j}) \nonumber \\
&-&g_{n}[\hat{a}_{n}^{\dagger}(f_n \hat{a}_{n}-h_{n}e^{\chi_{n}})+h_n-f_n \hat{a}_ne^{-\chi_{n}}].
\end{eqnarray}

\begin{figure}[!htb]
\begin{center}
\scalebox{0.67}[0.67]{\includegraphics{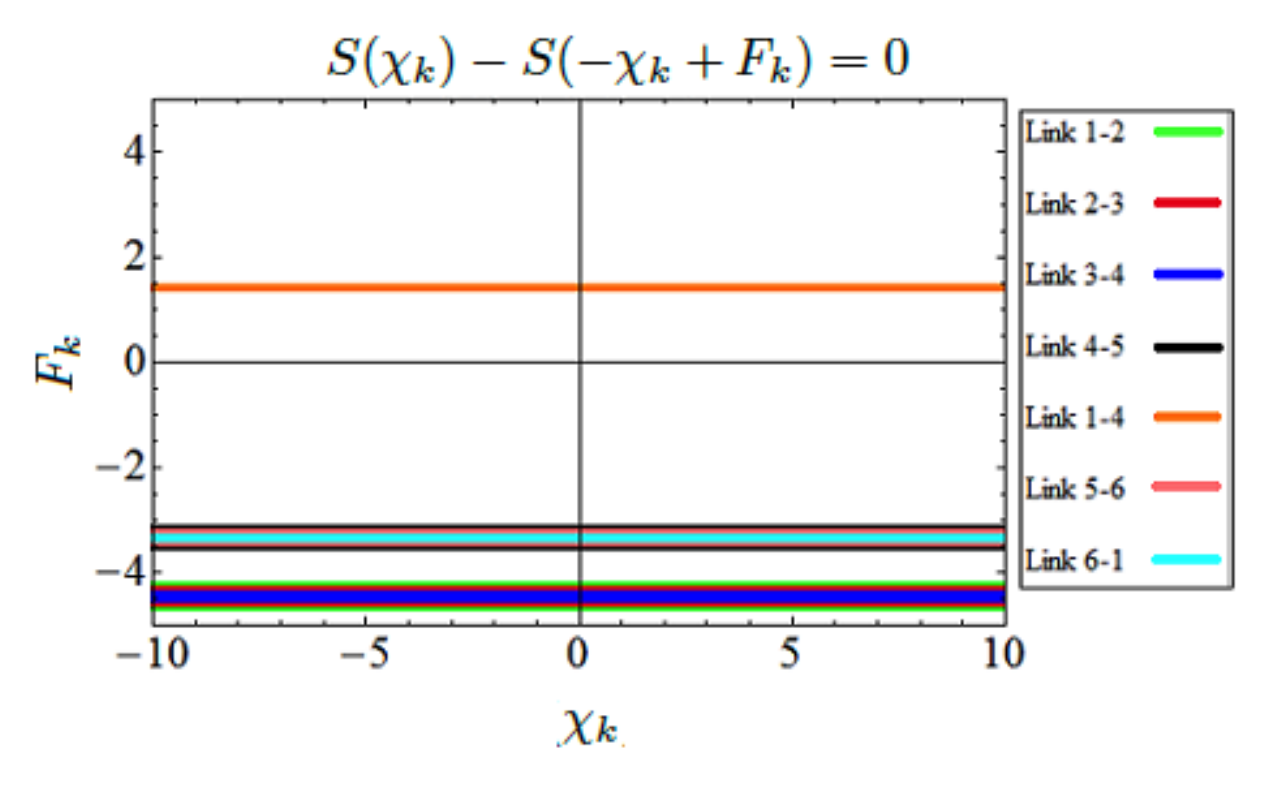}} \vspace{-3mm}
\hspace{1mm}   \caption{Numerical plots for the solution contours of $S(\chi)-S(-\chi+F)=0$ for all internal links in Fig.~\ref{Cyclic_main}(a). We consider parameters having the following values: $g_{12}=0.4557$,  $g_{21}=0.6282$, $g_{23}=0.97782$, $g_{32}=0.04859$, $g_{34}=0.9432$, $g_{43}=0.7787$, $g_{45}=0.9622$, $g_{54}=0.4298$,  $g_{56}=0.3023$, $g_{65}=0.3856$, $g_{61}=0.4667$, $g_{16}=0.06163$, $g_{41}=0.2779$, $g_{14}=0.09021$, $g_{1}=4.7019$, $h_{1}=8.7658$.} \label{cycle_contours1}
\end{center}
\end{figure}

To account for the lack of detailed balance condition we simply allow $g_{ij} \ne g_{ji}$. In Eq.~(\ref{master-operator-twisted_cycle}) the index "$n$" corresponds to the only node in the cyclic network that is connected to the reservoir. Similarly to the previous example, we obtain the following cumulant generating function for the case of the networks in Figs.~\ref{Cyclic_main}(a,b),
\begin{eqnarray}
S(\tau,{\bm \chi }) = g_{n}\bar{h}_{n}(e^{\chi_n}\zeta_{n}-1),
\label{bra-CGF_cycle}
\end{eqnarray}
where the average occupation number of the node $\zeta_{j}$ can be obtained by solving the  following set of linear equations,
\begin{eqnarray}
\label{eigenvalue-bra-coherent_cycle} \sum_{j}g_{jk}(e^{\chi_{jk}}\zeta_{j}-\zeta_{k})-\delta_{k,n}g_{k}(\zeta_{k}-e^{-\chi_{k}})=0.
\end{eqnarray}

\begin{figure}
\scalebox{0.7}[0.7]{\includegraphics{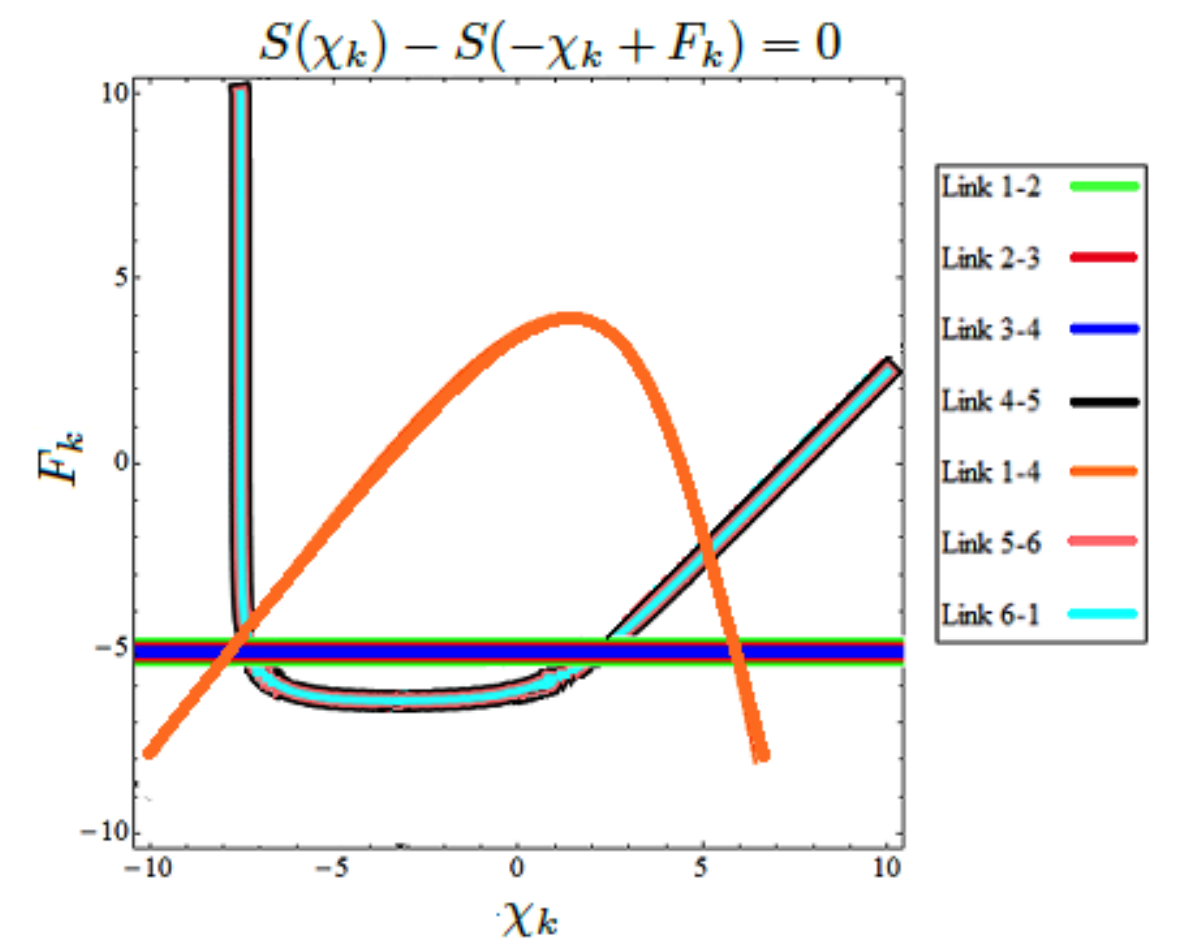}} 
\hspace{1mm}   \caption{Numerical plots for the solution contours of $S(\chi)-S(-\chi+F)=0$ for all internal links in Fig.~\ref{Cyclic_main}(b). The network parameters for this case are: $g_{12}=4.5654$,  $g_{21}=7.83$, $g_{23}=8.6823$, $g_{32}=0.6629$, $g_{34}=7.0427$, $g_{43}=7.4394$, $g_{45}=7.95$, $g_{54}=0.1431$,  $g_{56}=0.4052$, $g_{65}=2.5126$, $g_{61}=9.5782$, $g_{16}=0.08372$ ,$g_{41}=9.0737$, $g_{14}=0.81007$, $g_{2}=5.5299$, and $h_{2}=8.2477$.} \label{cycle_contours2}
\end{figure}
The numerical check for this case is done by inspecting the solution contours of  $S(\chi)-S(-\chi+F)=0$, as described in previous subsection. 

We consider two different cases: 

(a) when the reservoir is connected to the central node $1$, and

(b) when the reservoir is connected to a non-central node (e.g. the node $2$).   

We first consider case (a). 
 According to the plot (see Fig.~\ref{cycle_contours1}),  all the links have solution contours that are independent of $\chi$. Hence  the FRCC holds for all the links when the reservoir is attached to the three link junction $1$. The degeneracies in the values of $F$ are due to the charge conservation. Currents through the links  $(1,2)$,  $(2,3)$ and  $(3,4)$ are the same which leads to the same value of $F$. This degeneracy  is also seen for the links $(4,5)$,  $(5,6)$ and  $(6,1)$. Hence, in Fig.~\ref{cycle_contours1} we see only three sets of degenerate horizontal lines.
 
We perform similar numerical analysis for the case (b) when the reservoir is connected to node 2. The solution contours for $S(\chi)-S(-\chi+F)=0$ are shown in Fig.~\ref{cycle_contours2}. For this case we obtain degenerate constant lines of $F={\rm const}$ for the links  $(1,2)$,  $(2,3)$ and  $(3,4)$, thereby satisfying an FRCC. All the other links do not satisfy an FRCC.  

\section{Conclusion}
We demonstrated that FRCCs appear in fundamental models of nanoscale electric circuits. Many-body electron interactions, including Coulomb interactions and  exclusion interactions, do not break the FRCC prediction in the model of a chaotic cavity coupled to several leads.  This robustness can be used to extract information about relative sizes of single particle tunnelling barriers independently of electron interactions. To the best of our knowledge this is a unique example in which measurements of fluctuation relations can provide quantitative information that cannot be easily obtained by measuring lowest current cumulants at given voltages. We also demonstrated that the FRCCs extend to the quantum regime of coherence among electron trajectories. FRCCs can be exact even when generating functions cannot be derived. This reflects the fact that FRCCs follow from the 
properties of single particle geometric trajectories that separately show profound symmetries even when the complete stochastic evolution of a system is very complex.

An unusual property of FRCCs is that they are not directly related to the system's dissipation function. This distinguishes them from the vast number of previously found fluctuation relations for currents, entropy, heat and work.  
Instead, the fact that FRCCs follow from constraints on geometric trajectories relates them to the principle of Geometric Universality of Currents \cite{GUC} and the class of exact results in non-equilibrium physics called "no-pumping theorems" \cite{GUC,nopump}. On the other hand, similarity between FRCCs and standard fluctuation relations suggests that there can be a more fundamental theory in the background of both fluctuation and no-pumping relations. 
 
Our results should stimulate further theoretical and experimental studies of fluctuation relations,  including the search for unifying fundamental principles in non-equilibrium statistical mechanics and new measurement techniques that are enabled by FRCCs.

Recently, we have learned of another work that obtained a large class of FRCCs  \cite{mukameld-11} that do not directly relate to the dissipation functional of the total system.
Unlike our results, their FRCCs apply to a closed network topology without external particle reservoirs.

{\it Acknowledgment}. Authors thank D. Andrieux for useful discussion. The work at
LANL was carried out under 
DOE Contract No. DE-AC52-06NA25396. Work at NMC was supported  by NSF under ECCS-0925618.


\end{document}